\documentclass[twocolumn]{aastex631}
\usepackage{newtxtext,newtxmath}
\usepackage{amsmath}	

\usepackage{graphicx}%
\usepackage{amssymb,amsfonts}%
\usepackage{mathtools}

\newcommand{\frm}{\textit{Fermi/}}
\newcommand{\grb}{GRB 220426A }
\newcommand{\gr}{GRB 230812B }
\newcommand{\tnt}{$t_{90}\,$}
\newcommand{\wid}{$\mathcal{W}\,$}
\newcommand{\rat}{$\mathcal{H}\,$}
\newcommand{\alp}{$\alpha\,$}
\newcommand{\ind}{$\xi\,$}
\newcommand{\gam}{$\Gamma\,$}

\begin{document}

\title{Are Single-Zone Emission models Sufficient to Explain GRB 220426A and GRB 230812B?}

\correspondingauthor{Soumya Gupta, Sunder Sahayanathan}
\email{soumya.gupta1512@gmail.com}

\author[0000-0001-6621-259X]{Soumya Gupta$^{*}$}
\affiliation{Homi Bhabha National Institute, Mumbai, Maharashtra, India}
\affiliation{Bhabha Atomic Research Center, Mumbai, Maharashtra, India}

\author[0009-0009-8004-8314]{Sunder Sahayanathan$^{+}$}
\affil{Homi Bhabha National Institute, Mumbai, Maharashtra, India}
\affil{Bhabha Atomic Research Center, Mumbai, Maharashtra, India}

\author{Saharsh Shanu}
\affiliation{UM-DAE Centre for Excellence in Basic Sciences, Mumbai, Maharashtra, India}

\author{Rishabh Nath}
\affiliation{UM-DAE Centre for Excellence in Basic Sciences, Mumbai, Maharashtra, India}

\begin{abstract}
Gamma-ray bursts (GRBs) are the universe's most energetic phenomena (isotropic luminosity $\sim 10^{51} - 10^{54}$ ergs/s) lasting for a 
very short duration ($\sim$ milliseconds - a few seconds). Even after an average of one GRB detected per day, their emission mechanism remains contentious. Inferences drawn from the empirical modelling of the GRB spectrum are often inconclusive. Some studies favor the emission from a thermal blast of hot plasma, while others suggest a synchrotron emission originating from a rapid acceleration of particles at the expense of the burst energy. Under these scenarios, the spectral width of the burst (\wid), which is measured at half maxima, is expected to decrease with time. We show that for the GRB 220426A and GRB 230812B, \wid increases with time, raising serious concerns regarding the validity of these emission models. The results instead offer strong evidence that the GRB prompt phase involves the development of multiple emission zones, whose relative contributions change over time.
\end{abstract}


\section{Introduction}\label{into}
Despite decades of research, the prompt emission phase of Gamma-Ray Bursts (GRBs) is only partially understood. The complexity arises primarily from the sequence of energy conversion processes that result in the observed electromagnetic radiation \citep{meszaros93,rees_2005,beloborodov_2011}. Various theoretical models have been proposed to interpret this radiation under a thermal \citep{paczy_1986,piran_shemi_naray,peer_08, Gupta_2024} or non-thermal emission scenario \citep{Rees_internal,rees94, Tavani1996,cohen_1997,sari_1997,daigne_1998, Bonjak2009}. However, the GRB spectrum during the prompt phase is complex to be inferred under these processes alone, rather necessitating the inclusion of a hybrid emission mechanism \citep{Iyyani_etal_2016,zhang_2018,chen_21,gupta_230307}. Conversely, the emission processes are also routinely analyzed using an empirical function representing a broken power-law with a smooth exponential transition, commonly referred to as the band function \cite{band1993batse}. Particularly, the low energy spectral index, $\alpha$, of the band function is often used to validate the interpretation based on the synchrotron radiation by a relativistic electron distribution, accelerated in the outflow \citep{sari98,daigne_2011, Peer_2006,uhm_14,bbzhang_2016}. For instance, within the slow cooling synchrotron radiation limit, \alp should not be harder than $-2/3$, usually termed as the line of death (LOD) \citep{crider_1997, Preece_etal_1998}. This limit can be further curtailed to $-0.8$ when the spectral fit is performed using a combination of band and Planck function \citep{burgess_ryde_yu}. Under the fast cooling synchrotron process, the limiting value of \alp will be $-3/2$ \citep{sari_1997,daigne_2011}. Further, a hard spectrum, with an index of $-0.52$, can be obtained when the cooling process is associated with the decaying magnetic field \cite{bbzhang_2016}. Nevertheless, GRBs with typical $\alpha \sim -1$ cannot be perceived either under the synchrotron or thermal origin alone \citep{vurm_13,lundman2013,deng2014}. These results suggest that the limits on \alp can only provide a little insight into the origin of prompt emission of the GRB. In addition to these, the low photon statistics at the high-energy tail of the GRB spectra allow multiple models to reproduce the same spectrum, leading to interpretation ambiguities. 

Lately, a novel way to interpret the GRB spectrum was introduced by studying its spectral width at half maximum, $\mathcal{W}$, in $E^2 N(E)$ representation (in units of energy per unit surface per second). By analyzing the spectra of nearly 2000 GRBs, detected by \emph{BATSE} or \frm GBM, it was shown that the distribution of \wid peaked at $\sim 1$ \cite{axelsson_width}. This highlighted that most of the GRB spectra differ significantly from the narrow Planck function (\wid $=0.54$) and a broad non-thermal synchrotron emission. Recently, it was demonstrated that a composite thermal spectrum arising from a relativistically expanding fireball can be wider than the Planck function \cite{Gupta_2024}. The \wid under this scenario was related to the dynamics of the fireball, and its value $\sim 1$ corresponds to the matter-dominated phase of the burst. Despite these studies, the most probable \wid, obtained for the large sample of GRBs, can also be an artifact due to the post-processing of the unfolded data \cite{burgess_width}. Alternatively, a statistical fit to the GRB spectrum with \wid as a free parameter is capable of overcoming this limitation by providing confidence in the estimates.

The knowledge about the temporal evolution of \wid, instead of the time-integrated one, which further has the potential to identify the dominant emission mechanism, prevalent over the duration of the burst. It was also reported that the time-integrated analysis often falsely identifies the thermal component due to the curvature introduced by the spectral evolution \cite{burgess_ryde_2015}. Such analysis could mask the crucial spectral information when the emission is a combination of both thermal and non-thermal processes \cite{GillandGranot2021}. Moreover, the radiation resulting from an evolving non-thermal particle distribution under the synchrotron losses is found to explain the time-resolved spectrum of GRBs, though the time-integrated analysis rejected this scenario under LOD arguments \citep{Burgess_nat}. For the case of GRB 160821A, it was shown that the emission is unpolarised in the time-integrated regime, whereas a flip in polarization angle is observed in the time-resolved case \cite{Sharma_etal_2019}. These studies highlight the importance of time-resolved analysis to uncover the key components of the emission mechanism that remain hidden in time-integrated observation. 

In this work, we investigate the temporal evolution of \wid for the \frm GBM detected bursts, namely, \grb and GRB 230812B. The former was detected at 06:49:51 on 26 April 2022 \cite{220426_fermi} while the later on 12 August 2023 at 18:58:12. These GRBs are very bright with fluence greater than $10^{-4}$  \texttt{ergs/cm$^2$} \citep{22_fluence,23_fluence} which suffices the requirement for a statistically significant time-resolved spectral study \citep{yu_bright_fermi}. To facilitate the current study, we restructured the band function and the power-law with an exponential cutoff function (CPL) in terms of \wid and used it to fit the time-resolved spectra of these two GRBs. The evolution of best-fit parameters obtained was compared with the evolution of the thermal and non-thermal interpretation of the bursts. 
 
\section{Data analysis}\label{data}

To study the temporal evolution of the quantities defining the GRB spectrum during its prompt phase, we selected \grb and GRB 230812B, which are observed by \frm GBM. The observed data is reduced employing the standard software provided by \frm GBM instrumentation team. For each burst, \grb and \gr, the Time-Tagged Event (TTE) data from the two brightest Sodium Iodide (NaI) and the brightest bismuth germanate (BGO) detectors were analyzed (source viewing angle less than 60$^\circ$). A significant pile-up was observed in the case of GRB 230812B during the time windows T0+0.54--T0+1.70 s and T0+0.61--T0+1.12 s for the NaI and BGO detectors, respectively \citep{34694}, and hence they were omitted for the present analysis.

The \frm GBM light curve was extracted using the RMFIT software (version 4.3.2) for the brightest NaI detector, and time bins were created using the Bayesian block algorithm with a chance probability of $p_0 = 0.05$ \citep{scargle1998studies}. Thereafter, each time bin was required to achieve a detection significance greater than 8$\sigma$ above the background, estimated using a polynomial fit to the pre- and post-burst emission intervals. Bins that failed to meet this threshold were discarded entirely, as the associated spectral parameters would be unconstrained or dominated by systematic uncertainties in the background model. The selected bins were further required to contain a minimum of 800 net counts, defined as the difference between the total observed counts and the estimated background counts within the interval. This threshold was chosen to ensure adequate degrees of freedom for spectral fitting and to avoid parameter degeneracies in the photon model.  The bins satisfying both criteria simultaneously were retained as primary spectral intervals and carried forward without modification, preserving the original temporal resolution afforded by the Bayesian Block decomposition. In case of the bins that passed the significance threshold but fell below the 800-count minimum, an iterative merging procedure was applied. Here, each under-threshold bin was successively absorbed into its temporal neighbor with the lower net count total. This strategy will distribute photon statistics as uniformly as possible while maintaining strict temporal contiguity and avoiding the artificial broadening of individually bright intervals. Using this technique, 16 time-stamps for the \gr and 18 time-stamps for the \grb were obtained (Tables \ref{tab22} and \ref{tab23}). Guided by these results, spectra were generated using the \texttt{Make spectra} tool within the \texttt{gtburst} software from the Fermi Science Tools. The background was estimated from two time intervals, one preceding and one following the main GRB emission. 

The spectral fitting is performed using the Multi-Mission Maximum Likelihood framework, 3ML, which is an interface to Bayesian inference and likelihood calculations. For the present work, the Poisson-Gaussian likelihood is selected, which accounts for the Poisson data and a Gaussian background, whereas the posteriors were sampled using the DYNESTY algorithm \citep{dynesty}. Each spectral fit was performed using 700 live points. The 33-40 keV energy range was excluded from the analysis due to the presence of the iodine K-edge at 33.17 keV. The spectral fitting is performed using a restructured Band and CPL functions (Appendix \ref{band} and \ref{CPL}) where the free parameters for the former function are chosen to be $\alpha$, $E_p$, and \wid (equation \ref{eq_width}). Here, $E_p$ is the energy at which the spectrum peaks 
in $E^2 N(E)$ representation. For the CPL function, the free parameters are chosen to be \wid (equation \ref{eq4c}) and the energy where exponential roll-off, $E_c$. The function with a lower BIC, in addition to a well-constrained parameter posterior contour, was considered the best-fit model. The best-fit parameters for each time-resolved spectrum of \grb and \gr are given in Table \ref{tab22} and \ref{tab23}, respectively. This enabled us to obtain the temporal behavior of $\mathcal{W}$ and the other parameters (Figure \ref{fig_alp_ep} and \ref{fig_wd}), which can be readily compared with the evolution of the thermal and non-thermal interpretation of the bursts. 

\section{Spectral evolution of an expanding fireball}\label{thermal}

When the GRB is attributed to a thermal blast of plasma (fireball), the flow dynamics and evolution of the spectrum will fall primarily between two regimes, namely matter-dominated or radiation-dominated \citep{Goodman_86,paczy_1986,piran_shemi_naray}. In the latter case, the bulk Lorentz factor, \gam, of the flow will scale with the radius of the fireball beyond the photospheric radius, $R_{ph}$. Due to the dynamics of the flow, the observer will perceive the temperature to be constant and equal to the one at $R_{ph}$ \citep{piran_shemi_naray}. Hence, the spectrum in this case will be similar to the Planck function, which is narrower than the ones typically observed for the GRBs \citep{piran_review_fireball}. Contrary to this, in the matter-dominated regime, the energy of the blast is stored as the kinetic energy of the baryonic matter, and the expelled plasma coast with a constant velocity. The temperature during this phase decreases with the radius as a power law with an index $\sim 2/3$ \citep{piran_review_fireball,peer_08, Gupta_2024}. Generally, when the energy is shared between the radiation and the matter, the temperature will fall as $R^{-\xi}$, where $R$ is the instantaneous radius and $0\le\xi\le 2/3$. For a spherically symmetric burst, the light travel time effects will cause the observed spectrum at any instant to be a superposition of cold equatorial (on-axis) and hot high latitude (off-axis) emissions \citep{peer_08, Gupta_2024}. The shape of the composite instantaneous spectrum will depend upon $R_{ph}$, \gam, and \ind, while the photospheric temperature only shifts the spectrum on the photon energy scale \citep{Gupta_2024}. Additionally, the relativistic beaming effect will restrict the emission only from a surface that subtends a semi-vertical angle of 1/\gam at the center of the fireball \citep{Gupta_2024,uhm_zhang_curvature}. Hence, for \gam $\gg 1$ the composite instantaneous spectrum will be marginally broader than the Planck function \citep{Gupta_2024}.

\begin{figure}
    \centering
    \includegraphics[width = 0.35\textwidth]{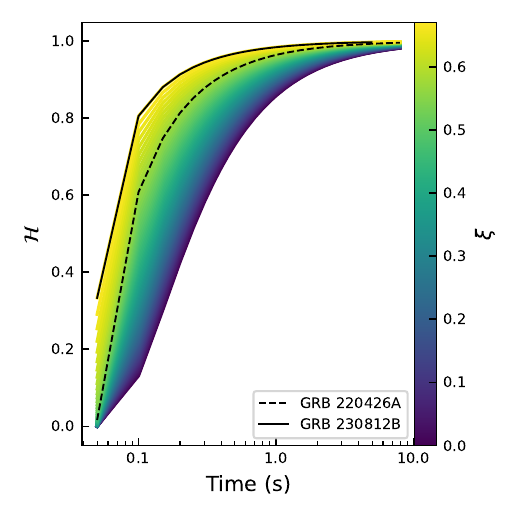}
    
	\caption{ Under the fireball scenario, the change in \rat with expansion 
    is shown for different $\xi$. The black dashed line depicts the trend for 
    \grb where \tnt is 8s and \ind is 0.57. On the other hand, the black solid 
    line represents the evolution in the case of \gr where \tnt is 5s and \ind is 0.67.} 
    \label{fig_flx}
\end{figure}

In case of the time-resolved spectrum obtained during a finite interval within the burst duration, the fireball would have expanded significantly in space, and the spatially integrated instantaneous spectrum will be much broader than the Planck function \citep{peer_08,ruffini_13, Gupta_2024}. For instance, let us assume the fireball expands from an initial radius of $R_0$ to $R$ during an interval $\Delta t$. The shape of the time-resolved spectrum and \wid will then be governed by the relative strength of the instantaneous spectra at $R_0$ and $R$. The ratio of the peak emitted power at these radii will vary as \rat $\sim (R_0/R)^{3-4\xi}$ and \wid decreases with the increasing \rat (Appendix \ref{fireball}). Due to the limiting condition on $\xi$ ($\lesssim 2/3$), the peak flux at $R_0$ will be less than that of $R$. For a given set of initial burst parameters, this limit decides the maximum \wid of the time-resolved spectrum, which approaches that of a Planck function as $\xi \to 0$. With the evolution of time, $R$ increases and for the same interval $\Delta t$, \rat increases rapidly as $\sim(1-2 c \beta \Gamma^2 \Delta t/R)^{3-4\xi}$ (Figure \ref{fig_flx}) and \wid approaches the width of the instantaneous spectrum. The peak energy ($E_p$) of the time-resolved spectrum will be governed by the temperature of the fireball at $R$. As the fireball evolves, the temperature falls as $\sim R^{-\xi}$ and consistently, $E_p$ decreases with time following the index $\xi$ (Appendix \ref{fireball}). 

Our analysis of GRBs 220426A and 230812B suggested that the evolution of $E_p$ obtained from the spectral fit largely depicted a decreasing trend, while the fall was rather monotonic in the case of the latter (Figure \ref{fig_alp_ep}). 
Typically, the fall in $E_p$ can be approximated as a power-law in time with an index $\sim -1.2$, and such a rapid change in $E_p$ cannot be attained under an evolving fireball (unless $\xi > 2/3$). 

\section{Spectral evolution under synchrotron losses}\label{nonthermal}

The GRB prompt emission, alternatively, can be non-thermal in nature, associated with the synchrotron radiative process. To mimic this, we consider a scenario where a relativistic Maxwellian distribution of electrons, $\gamma^2\exp(-2 \gamma/\gamma_p)$, loses their energy through the synchrotron process. Here, $\gamma$ is the electron Lorentz factor and $\gamma_p$ corresponds to the peak of the distribution. The spectral evolution can be obtained by convolving the instantaneous electron distribution (Appendix \ref{synch}) with the single particle emissivity. The main parameters that govern the spectral shape and its evolution are \gam, $\gamma_p$, and the magnetic field $B$. From the synchrotron theory, the initial peak energy of the spectrum $E_p$ can be expressed in terms of these parameters as $\mathcal{C}\,\Gamma\,\gamma_p^2 B$ (where $\mathcal{C}\approx1.144\, eh/mc$) \citep{rybicki}, and the energy loss time-scale of the electrons with Lorentz factor $\gamma_p$ will be $\mathcal{D}\, \Gamma\gamma_p^{-1}B^{-2}$ (Appendix \ref{synch}). Hence, for a given $\Gamma$, the parameters $\gamma_p$ and $B$ can be constrained from the observed spectral peak and the duration of the burst. In the case of GRB 230812B, the initial spectral peak and burst duration are $\sim 2400$ keV and $\sim 5$ s ($T_{90}$), which results in $\gamma_p$ and $B$ as $\sim 1.4\times10^5$ and $\sim 1.44$ G, respectively, for the choice of $\Gamma$ as 1000. Similarly, these parameters are found to be $\gamma_p=5.4\times10^4$ and $B=1.57$ G for \grb with initial spectral peak and $T_{90}$ as $\sim 385$ keV and $\sim 7.6$ s. Since the parameters are chosen to reproduce these observed quantities, the choice of $\Gamma$ will affect only the magnitude of these estimated parameters and will not modify the spectral shape of the GRB or its evolution.

\begin{figure}
    \centering
    \includegraphics[width = 0.36\textwidth]{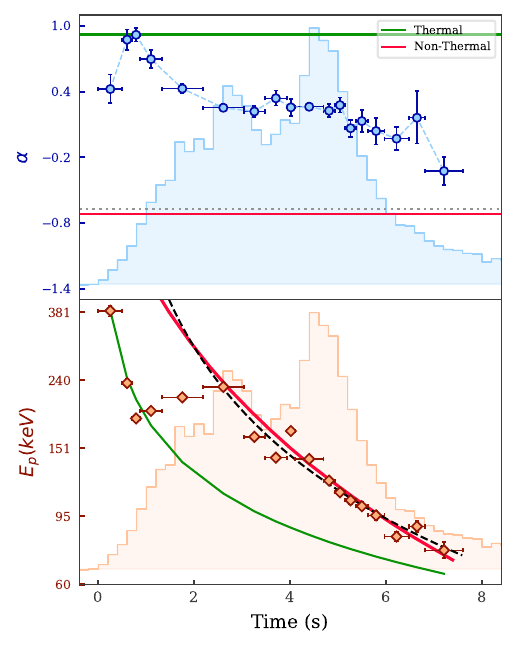}
    \includegraphics[width = 0.36\textwidth]{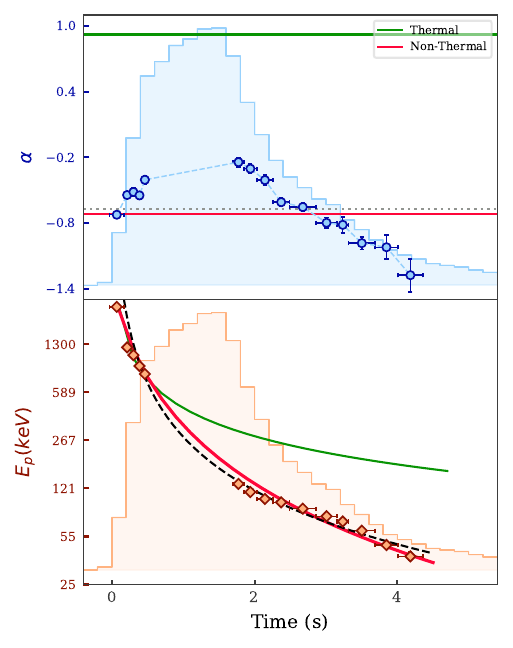}
    
	\caption{This figure depicts the evolution of the best-fit spectral parameter for Fermi/GBM observations GRB 220426A (top) and \gr (bottom). In each plot, the upper and lower panels represent the evolution of the low-energy index $\alpha$ (blue circles) and the variation in peak energy $E_p$ (dark yellow diamonds), respectively. The green solid line in the upper panel is the slope of the low-energy index of the Planck function ($\sim 1$). The red solid line in the same panel is the slope of the low-energy index of the synchrotron spectrum due to the Maxwellian particle distribution. The grey dotted line in the upper panels of each plot marks the LOD. The green solid line in the lower panel represents the evolution of $E_p$ under the fireball scenario, whereas the evolution of synchrotron emission $E_p$ is represented in red. The photospheric radius of the expanding fireball in the case of \grb is $\sim 10^{13}$ cm and $10^{14}$ cm in the case of GRB 230812B. The black dashed line in the bottom panel of each figure portrays the power law fit to the data.}
    \label{fig_alp_ep}
\end{figure}

The evolution of the Maxwellian electron distribution under synchrotron loss will shift $E_p$ to lower energies, and this is shown in Figure \ref{fig_alp_ep} along with the observed values. This scenario was able to explain the evolution of $E_p$ in the case of \gr reasonably well; whereas, for GRB 220426A, the initial evolution of $E_p$ ($< 4$ sec) does not agree with the observations. Though the evolution of $E_p$ supports a synchrotron origin of the burst (at least in the case of GRB 230812B), variation in \alp portrays a contradictory picture. Throughout the burst duration, the \alp of \grb is harder than LOD, thereby disfavoring the synchrotron origin. On the other hand, for GRB 230812B, the initially hard \alp softens during the decay phase, crossing the LOD at $\sim 3 s$. We may argue this as a result of the emission process transiting from thermal to non-thermal origin \citep{gupta_230307}; however, the rate of decrease in $E_p$ is clearly in disagreement with the thermal interpretation. 

\begin{figure}
    \includegraphics[width = 0.5\textwidth]{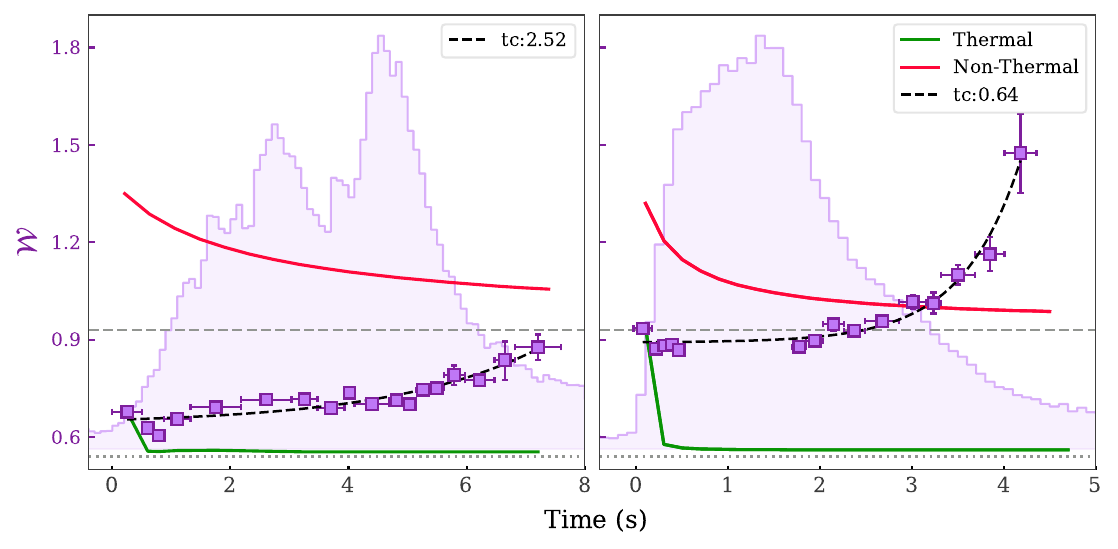}
	\caption{In this figure, the temporal evolution of \wid and comparison with thermal and non-thermal emission scenarios is shown. The evolution of \wid (purple squares) for the case of \grb is shown in the left and on the right for GRB 230812B. The grey dotted line represents the \wid of the Planck function and the grey dashed line corresponds to the \wid of the synchrotron spectrum obtained from a mono-energetic electron distribution. The green solid line in both figures indicates the evolution of spectral broadness under the expanding fireball scenario. The red solid line portrays the evolution of \wid of the synchrotron spectrum from the Maxwellian distribution of particles. The black-dashed line in each panel depicts the best-fit exponential 
    function with characteristic evolution time, tc.} 
    \label{fig_wd}
\end{figure}

\section{Evolution of the spectral width (\wid)}\label{specwid}

The spectral fit using the restructured band function allowed us to study the evolution of \wid, which was never attempted earlier and could possibly provide a better understanding of the prompt phase of the GRB. Under the thermal scenario, the relativistic expansion of the plasma reduces the \wid of the composite spectrum rapidly to the instantaneous spectral width (Figure \ref{fig_wd}). On the other hand, when the GRB is attributed to the synchrotron process, the \wid of the spectrum during the initial phase of the burst will be $\sim 1.39$, consistent with the Maxwellian electron distribution. Further evolution of the spectrum is governed by the synchrotron loss rate, which varies quadratically with the particle energy. Hence, the high-energy electrons lose their energy efficiently and get depleted faster with time, and this causes the particle distribution to narrow down along with the \wid of the emitted synchrotron spectrum to decrease as well. Therefore, irrespective of the thermal or non-thermal origin, the \wid is found to be decreasing with time (Figure \ref{fig_wd}).

On the contrary, the evolution of \wid in the case of \gr and \grb is observed to be increasing, suggesting a more complex emission scenario responsible for these bursts. The increase is prominent in \gr where \wid rises from $\sim 0.9$ to $\sim 1.6$; while in GRB 220426A, the \wid remains relatively constant around $\sim 0.7$, with a moderate increase during the decaying phase of the second peak (Figure \ref{fig_wd} and Table \ref{tab22}). Notably, the spectra of both GRBs remain broader than that of a Planck function (\wid  $\approx 0.54$) throughout their durations. Furthermore, the spectrum of \gr evolves from an initial width which is close to the synchrotron single particle emissivity (\wid $\approx 0.93$)\citep{axelsson_width} to a much broader spectrum (\wid $\approx 1.63$); whereas, the \wid in \grb remains less than \wid $=0.93$ throughout the burst. 

\section{Discussion and conclusion}\label{conclusion}

The temporal behavior of the \grb and \gr presented in this work reflects the complexity involved in the emission process. The evolution of \alp shown in Figure \ref{fig_alp_ep} suggests that the emission cannot be interpreted as synchrotron emission (throughout the GRB 220426A and during the initial phase of GRB 230812B). On the contrary, the evolution of $E_p$ supports the synchrotron origin (during the later phase of \grb and throughout GRB 230812B). The evolution of \wid, on the other hand, disfavors both these scenarios (Figure \ref{fig_wd}). These results highlight the drawback of inferring the selected bursts' prompt phase under an idealized emission scenario. Therefore, this study indicates that the burst could possibly be associated with multiple emission zones, which can effectively widen the observed spectrum during evolution. The decrease in $E_p$ also hints that the dominant emission shifts between the zones with time.
 
\begin{figure}
            \centering
            \includegraphics[width=0.5\textwidth]{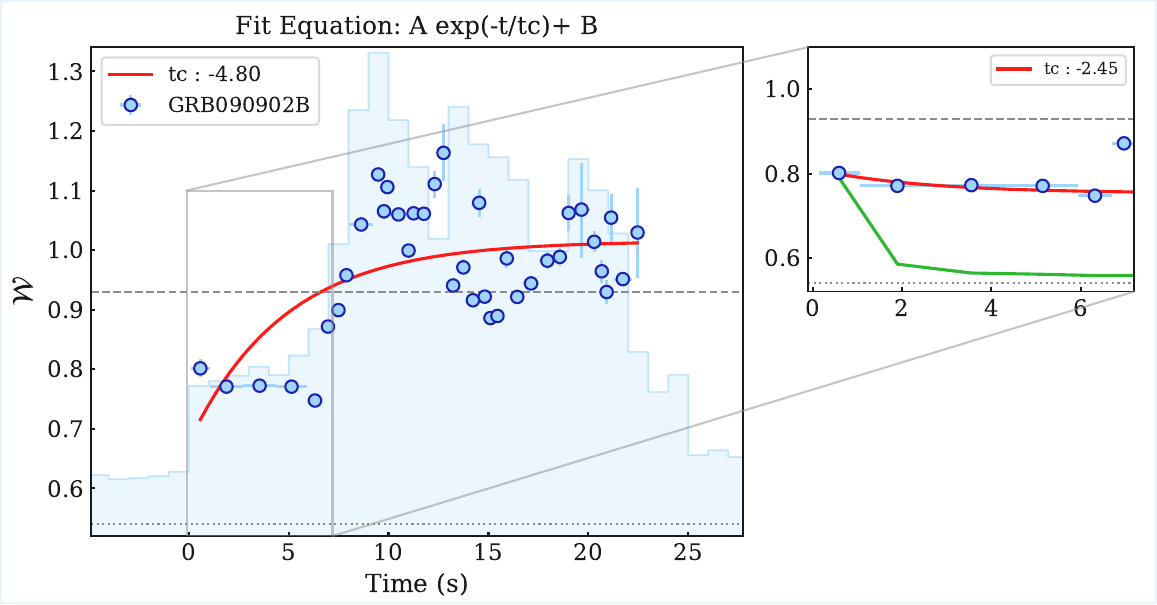}
            \caption{The figure represents the evolution of \wid for the case of GRB 090902B. The red line depicts the best-fit exponential function. The grey-dashed and grey-dotted line represents the \wid of the synchrotron spectrum obtained from the mono-energetic electron distribution, and the Planck function, respectively. The zoomed-in plot on the right depicts the trend of \wid during the first 8 s. The red line depicts the best-fit exponential function during the initial phase. The green line represents the evolution of \wid for the expanding fireball scenario. }
            \label{fig_090902}
        \end{figure}

The increasing \wid over the prompt duration can be specific for these selected GRBs. On the contrary, for the GRBs whose emission mechanism is well constrained from multiple studies, the evolution of \wid may be consistent. As a test case, we repeat the current study on a multi-peak GRB 090902B 	whose emission mechanism during the initial phase is known to be thermal dominant through different studies \citep{peer090902B,zhang_090902,mizuta_090902,Bromberg_090902}. However, the evolution of \wid for this GRB over the entire prompt phase is found to be chaotic, as shown in Figure \ref{fig_090902}. Though during the initial phase (T0-T0+7s), \wid shows a gradual decline from $\sim 0.8$ to $\sim 0.77$, it does not reflect the trend shown by a thermal evolution from an expanding fireball (inset in Figure \ref{fig_090902}). The \wid increases rapidly thereafter till T0+9.5s, followed by a decline in correlation with the pulse profile. We do not see any appreciable correlation of \wid with the pulse profile after this dominant peak. This complex behavior of \wid again indicates the dominance of a probable multi-zone emission scenario which cannot be interpreted under simple thermal/non-thermal models.

One possible scenario where such a multiple-emission zone can be achieved is through internal shocks formed by a sporadically active central engine. Under this interpretation, the faster-moving ejecta collide with the slower-moving ones, and the kinetic energy lost in the collision is transferred to the constituent particles through internal shocks \citep{Rees_internal,spada2001}. The decrease in $E_p$ possibly indicates the gradual decline in the activity of the central engine. The broadening of the time-resolved spectrum hints at an increase in the number of internal shocks formed during the evolution. This inference can also provide a direct explanation for the 
multi-pulse GRB \citep{kobayashi_1997,daigne_1998}. However, in this work we show \wid $< 0.93$ for the entire duration of \grb and the initial phase of \gr. This \wid may be hard to perceive under the synchrotron emission initiated by the particles accelerated at internal shocks.

Alternatively, multiple emission zones can also arise due to a rapid growth of turbulent plasma, such that the fluid becomes increasingly chaotic. This turbulent environment can induce multiple magnetic reconnection events, resulting in the formation of several distinct emission zones, each forming its own population of electrons. The cumulative effect of radiation from these multiple zones can broaden the observed spectrum. Additionally, the decrease in the spectral peak may reflect a redistribution of energy content among the growing number of emission regions. This scenario of a turbulent, reconnection-driven emission zone, leading to spectral broadening and energy redistribution, can also be visualized under the Internal-Collision-induced MAgnetic Reconnection 
and Turbulence (ICMART) model \citep{zhang_icmart}.

The complexity in the emission mechanism during the prompt phase of the \grb and \gr is evident from the temporal analysis of the spectral width, which is highlighted in this work. A broader perspective on the details of the emission process can be achieved by including the polarization property of the GRB prompt phase. This would provide an intricate understanding of the realignment of magnetic fields during the burst, which can be comprehended by studying the variation in the degree of polarization and the polarization angle. Upcoming missions like COSI and POLAR2 have the potential to perform such a study, leading to a better understanding of this enigmatic phenomenon.

{The authors acknowledge the referee for the valuable suggestions. This research has used data obtained through the HEASARC Online Service, provided by the NASA-GSFC, in support of NASA High Energy 
Astrophysics Programs.}
\begin{table*}
\centering
\caption{Time-Resolved spectral analysis parameter details  of \grb\,using \frm GBM data}
\label{tab22}
\begin{tabular}{|c|c|c|c|c|c|c|c|c|c|}
    \hline
    $(T_{i},T_{f})$ &  $\alpha$ & $\mathcal{W}_{Band}$ & $ E_p$ & $BIC_{Band}$ & $\mathcal{W}_{CPL} $ & $ E_c$ & $BIC_{CPL}$ & Flux (8-900 keV) & Best Fit Model\\
    $sec$ & & & $keV$ & & & keV & & $\times 10^{-5}\, ergs/cm^2/s$ & \\[3pt]
    \hline    
    \hline
    $(0.00,0.50)$& $0.42^{0.13}_{-0.13}$ & $1.64^{0.77}_{-0.78}$ & $357.39^{19.47}_{-19.99}$ & 772.66 & $0.68^{0.02}_{-0.02}$ & $158.87^{13.87}_{-14.60}$ & 754.47 & $0.72^{0.04}_{-0.04}$ & CPL\\
    $(0.50,0.70)$& $0.87^{0.09}_{-0.10}$ & $1.59^{0.84}_{-0.86}$ & $202.48^{7.68}_{-7.73}$ & -146.38 & $0.63^{0.01}_{-0.02}$ & $81.98^{6.51}_{-6.80}$ & -170.61 & $1.04^{0.05}_{-0.05}$ & CPL\\
    $(0.70,0.87)$& $0.92^{0.06}_{-0.06}$ & $1.64^{0.79}_{-0.76}$ & $180.95^{5.00}_{-5.13}$ & -275.00 & $0.61^{0.01}_{-0.01}$ & $63.48^{4.17}_{-4.33}$ & -307.82 & $1.13^{0.04}_{-0.04}$ & CPL\\
    $(0.87,1.32)$& $0.70^{0.08}_{-0.08}$ & $0.67^{0.02}_{-0.02}$ & $165.60^{4.89}_{-4.86}$ & 853.95 & $0.66^{0.01}_{-0.01}$ & $72.34^{2.69}_{-2.58}$ & 841.99 & $1.42^{0.05}_{-0.04}$ & Band\\
    $(1.32,2.18)$& $0.43^{0.04}_{-0.04}$ & $0.72^{0.01}_{-0.01}$ & $179.58^{3.95}_{-3.88}$ & 1646.96 & $0.69^{0.00}_{-0.00}$ & $88.07^{1.97}_{-2.00}$ & 1666.19 & $2.16^{0.04}_{-0.04}$ & Band\\
    $(2.18,3.03)$& $0.25^{0.03}_{-0.03}$ & $0.74^{0.01}_{-0.01}$ & $195.36^{3.93}_{-3.84}$ & 1736.70 & $0.72^{0.00}_{-0.00}$ & $101.93^{1.94}_{-1.92}$ & 1779.23 & $3.02^{0.05}_{-0.05}$ & Band\\
    $(3.03,3.48)$& $0.22^{0.05}_{-0.05}$ & $0.72^{0.01}_{-0.01}$ & $146.74^{3.03}_{-3.15}$ & 834.26 & $0.72^{0.01}_{-0.01}$ & $73.71^{2.48}_{-2.42}$ & 833.52 & $1.64^{0.04}_{-0.03}$ & Band\\
    $(3.48,3.93)$& $0.34^{0.07}_{-0.07}$ & $0.82^{-0.12}_{-0.14}$ & $133.21^{3.95}_{-4.68}$ & 912.56 & $0.69^{0.01}_{-0.01}$ & $60.70^{1.87}_{-1.85}$ & 904.88 & $1.68^{0.09}_{-0.04}$ & Band\\
    $(3.93,4.10)$& $0.26^{0.08}_{-0.08}$ & $0.77^{0.03}_{-0.03}$ & $130.19^{4.42}_{-4.38}$ & -154.02 & $0.74^{0.01}_{-0.01}$ & $75.41^{3.66}_{-3.62}$ & -134.28 & $2.30^{0.08}_{-0.07}$ & Band\\
    $(4.10,4.69)$& $0.26^{0.03}_{-0.03}$ & $0.70^{0.01}_{-0.01}$ & $126.73^{1.52}_{-1.73}$ & 1287.02 & $0.71^{0.00}_{-0.00}$ & $62.19^{1.33}_{-1.31}$ & 1297.69 & $2.56^{0.03}_{-0.03}$ & Band\\
    $(4.69,4.93)$& $0.23^{0.06}_{-0.06}$ & $1.63^{0.80}_{-0.79}$ & $110.36^{2.24}_{-2.21}$ & 220.83 & $0.71^{0.01}_{-0.01}$ & $54.58^{2.04}_{-2.06}$ & 214.18 & $1.91^{0.03}_{-0.03}$ & CPL\\
    $(4.93,5.15)$& $0.28^{0.07}_{-0.07}$ & $1.68^{0.78}_{-0.81}$ & $105.01^{2.35}_{-2.32}$ & -19.39 & $0.70^{0.01}_{-0.01}$ & $49.28^{2.17}_{-2.17}$ & -29.94 & $1.42^{0.03}_{-0.03}$ & CPL\\
    $(5.15,5.37)$& $0.07^{0.08}_{-0.08}$ & $1.63^{0.81}_{-0.78}$ & $95.01^{2.49}_{-2.55}$ & 66.34 & $0.75^{0.01}_{-0.01}$ & $51.45^{2.85}_{-2.80}$ & 62.94 & $1.06^{0.02}_{-0.02}$ & CPL\\
    $(5.37,5.62)$& $0.13^{0.10}_{-0.10}$ & $1.55^{0.83}_{-0.80}$ & $83.94^{2.90}_{-2.83}$ & 59.57 & $0.75^{0.02}_{-0.02}$ & $47.90^{2.92}_{-2.84}$ & 57.75 & $0.82^{0.02}_{-0.02}$ & CPL\\
    $(5.62,5.97)$& $0.04^{0.12}_{-0.12}$ & $0.79^{0.03}_{-0.03}$ & $77.07^{3.23}_{-3.32}$ & 406.62 & $0.78^{0.02}_{-0.02}$ & $47.02^{2.77}_{-2.79}$ & 411.08 & $0.57^{0.03}_{-0.02}$ & Band\\
    $(5.97,6.47)$& $-0.03^{0.11}_{-0.11}$ & $1.62^{0.79}_{-0.79}$ & $71.14^{2.19}_{-2.15}$ & 768.64 & $0.78^{0.02}_{-0.02}$ & $42.16^{2.72}_{-2.67}$ & 770.26 & $0.40^{0.01}_{-0.01}$ & CPL\\
    $(6.47,6.82)$& $0.16^{0.24}_{-0.24}$ & $0.84^{0.06}_{-0.06}$ & $59.31^{3.86}_{-4.06}$ & 272.44 & $0.81^{0.02}_{-0.02}$ & $41.15^{3.34}_{-3.20}$ & 274.72 & $0.30^{0.02}_{-0.02}$ & Band\\
    $(6.82,7.60)$& $-0.33^{0.13}_{-0.13}$ & $0.88^{0.04}_{-0.04}$ & $59.08^{2.55}_{-2.52}$ & 1162.03 & $0.88^{0.03}_{-0.03}$ & $45.21^{4.18}_{-4.01}$ & 1175.63 & $0.22^{0.01}_{-0.01}$ & Band\\
        
    \hline
\end{tabular}

\end{table*}

\begin{table*}
\centering
\caption{Time-Resolved spectral analysis parameter details of \gr\,using \frm GBM data}
\label{tab23}  
\begin{tabular}{|c|c|c|c|c|c|c|c|c|c|}
    \hline
    $(T_{i},T_{f})$ &  $\alpha$ & $\mathcal{W}_{Band}$ & $ E_p$ & $BIC_{Band}$ & $\mathcal{W}_{CPL} $ & $ E_c$ & $BIC_{CPL}$ & Flux (8-900 keV) & Best Fit\\
    $sec$ & & & $keV$ & & & keV & & $\times 10^{-5}\, ergs/cm^2/s$ & Model  \\[3pt]
    \hline
    \hline
    $(-0.03,0.17)$& $-0.72^{0.02}_{-0.02}$ & $1.66^{0.79}_{-0.78}$ & $2316.84^{121.39}_{-122.59}$ & 405.53 & $0.93^{0.01}_{-0.01}$ & $1888.70^{116.83}_{-115.33}$ & 425.84 & $9.22^{0.32}_{-0.33}$ & CPL\\
    $(0.17,0.26)$& $-0.54^{0.03}_{-0.03}$ & $1.61^{0.79}_{-0.77}$ & $1192.74^{51.16}_{-48.77}$ & -552.01 & $0.87^{0.01}_{-0.01}$ & $849.90^{48.08}_{-47.69}$ & -535.87 & $16.04^{0.49}_{-0.48}$ & CPL\\
    $(0.26,0.35)$& $-0.50^{0.03}_{-0.03}$ & $0.96^{0.5}_{-0.5}$ & $870.69^{50.06}_{-48.48}$ & -317.63 & $0.88^{0.01}_{-0.01}$ & $734.25^{38.26}_{-36.73}$ & -287.20 & $17.97^{0.48}_{-0.45}$ & CPL\\
    $(0.35,0.43)$& $-0.53^{0.03}_{-0.03}$ & $0.92^{0.3}_{-0.3}$ & $782.69^{37.62}_{-38.16}$ & -508.94 & $0.88^{0.01}_{-0.01}$ & $626.92^{29.74}_{-28.92}$ & -484.14 & $19.44^{0.54}_{-0.51}$ & CPL\\
    $(0.43,0.50)$& $-0.40^{0.04}_{-0.03}$ & $1.05^{0.3}_{-0.3}$ & $569.13^{27.57}_{-28.20}$ & -514.67 & $0.87^{0.01}_{-0.01}$ & $500.34^{24.56}_{-24.81}$ & -443.87 & $20.24^{0.54}_{-0.50}$ & CPL\\
    $(1.70,1.85)$& $-0.24^{0.04}_{-0.04}$ & $0.88^{0.02}_{-0.02}$ & $130.90^{3.23}_{-3.21}$ & 108.52 & $0.85^{0.01}_{-0.01}$ & $105.76^{3.25}_{-3.21}$ & 229.14 & $6.06^{0.11}_{-0.10}$ & Band\\
    $(1.85,2.03)$& $-0.30^{0.04}_{-0.04}$ & $0.90^{0.02}_{-0.02}$ & $114.44^{2.59}_{-2.65}$ & 301.49 & $0.87^{0.01}_{-0.01}$ & $95.76^{3.12}_{-2.99}$ & 447.46 & $4.56^{0.09}_{-0.09}$ & Band\\
    $(2.03,2.26)$& $-0.41^{0.04}_{-0.04}$ & $0.95^{0.02}_{-0.02}$ & $102.17^{2.51}_{-2.52}$ & 489.20 & $0.92^{0.01}_{-0.01}$ & $97.54^{3.43}_{-3.49}$ & 633.08 & $3.54^{0.07}_{-0.07}$ & Band\\
    $(2.26,2.49)$& $-0.61^{0.04}_{-0.04}$ & $0.93^{0.02}_{-0.02}$ & $96.61^{2.26}_{-2.25}$ & 381.00 & $0.94^{0.01}_{-0.01}$ & $84.90^{3.19}_{-3.32}$ & 444.54 & $2.54^{0.05}_{-0.04}$ & Band\\
    $(2.49,2.86)$& $-0.65^{0.04}_{-0.04}$ & $0.96^{0.01}_{-0.01}$ & $87.07^{1.88}_{-1.89}$ & 1104.59 & $0.98^{0.01}_{-0.01}$ & $83.77^{2.95}_{-2.99}$ & 1199.02 & $2.15^{0.04}_{-0.03}$ & Band\\
    $(2.86,3.15)$& $-0.80^{0.05}_{-0.05}$ & $1.02^{0.02}_{-0.02}$ & $76.83^{2.11}_{-2.05}$ & 620.02 & $1.04^{0.02}_{-0.02}$ & $83.91^{4.13}_{-4.12}$ & 702.24 & $1.72^{0.03}_{-0.03}$ & Band\\
    $(3.15,3.32)$& $-0.82^{0.07}_{-0.07}$ & $1.01^{0.03}_{-0.03}$ & $70.48^{2.41}_{-2.40}$ & -353.30 & $1.04^{0.03}_{-0.03}$ & $74.39^{5.37}_{-5.50}$ & -308.96 & $1.36^{0.04}_{-0.04}$ & Band\\
    $(3.32,3.69)$& $-0.98^{0.06}_{-0.06}$ & $1.10^{0.03}_{-0.03}$ & $60.52^{1.63}_{-1.67}$ & 764.88 & $1.16^{0.03}_{-0.03}$ & $79.31^{4.99}_{-4.79}$ & 829.11 & $1.08^{0.02}_{-0.02}$ & Band\\
    $(3.69,4.01)$& $-1.02^{0.11}_{-0.11}$ & $1.16^{0.05}_{-0.05}$ & $47.87^{2.21}_{-2.13}$ & 484.62 & $1.29^{0.06}_{-0.06}$ & $77.56^{6.81}_{-6.80}$ & 539.37 & $0.74^{0.02}_{-0.02}$ & Band\\
    $(4.01,4.36)$& $-1.28^{0.15}_{-0.15}$ & $1.48^{0.12}_{-0.12}$ & $39.70^{2.77}_{-2.64}$ & 450.14 & $1.81^{0.18}_{-0.18}$ & $116.95^{17.63}_{-17.55}$ & 510.79 & $0.60^{0.03}_{-0.03}$ & Band\\
    $(4.36,4.92)$& $-1.25^{0.15}_{-0.16}$ & $1.52^{0.12}_{-0.12}$ & $36.03^{2.49}_{-2.48}$ & 1109.95 & $2.14^{0.28}_{-0.26}$ & $138.46^{22.86}_{-22.67}$ & 1184.90 & $0.46^{0.02}_{-0.02}$ & Band\\
    
    \hline
\end{tabular}

\end{table*}

\begin{figure}
    \centering
    \includegraphics[width = 0.5\textwidth]{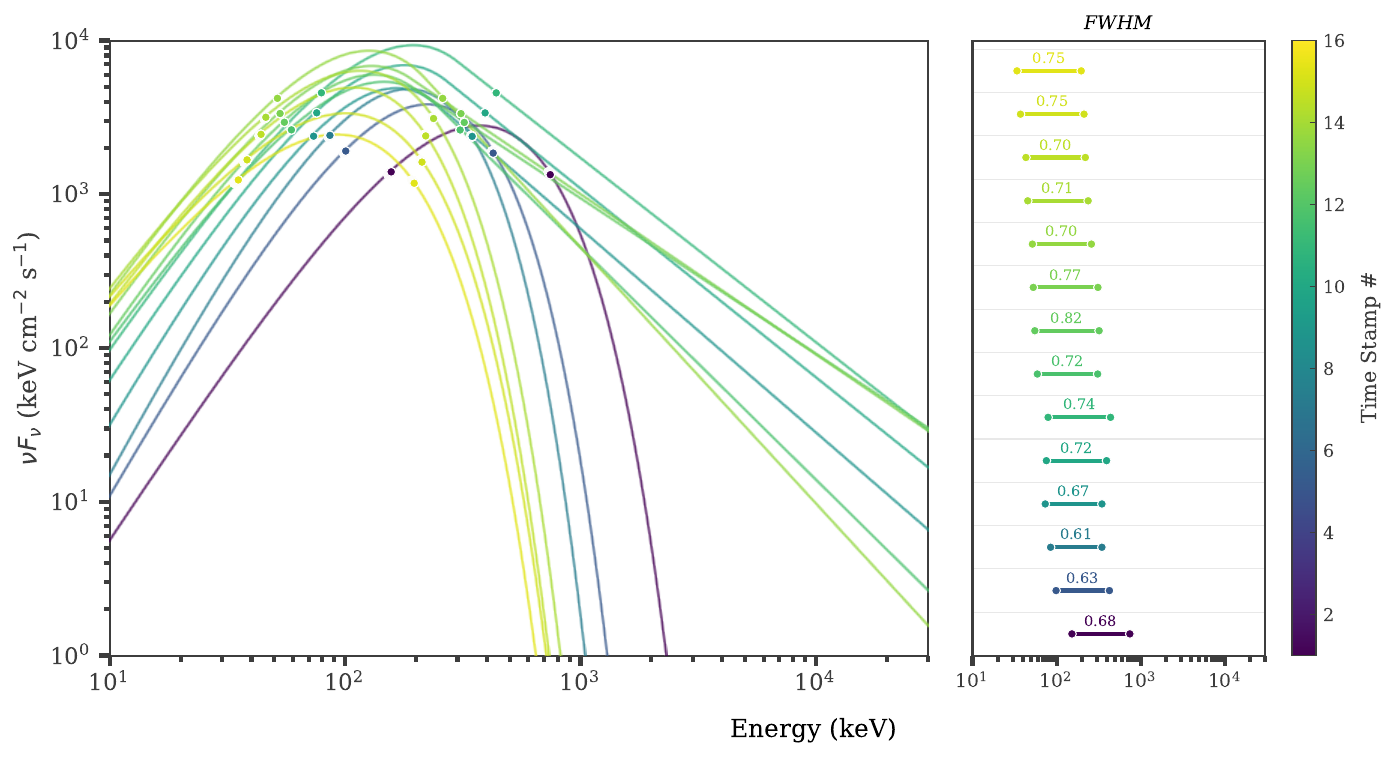}
    \includegraphics[width = 0.5\textwidth]{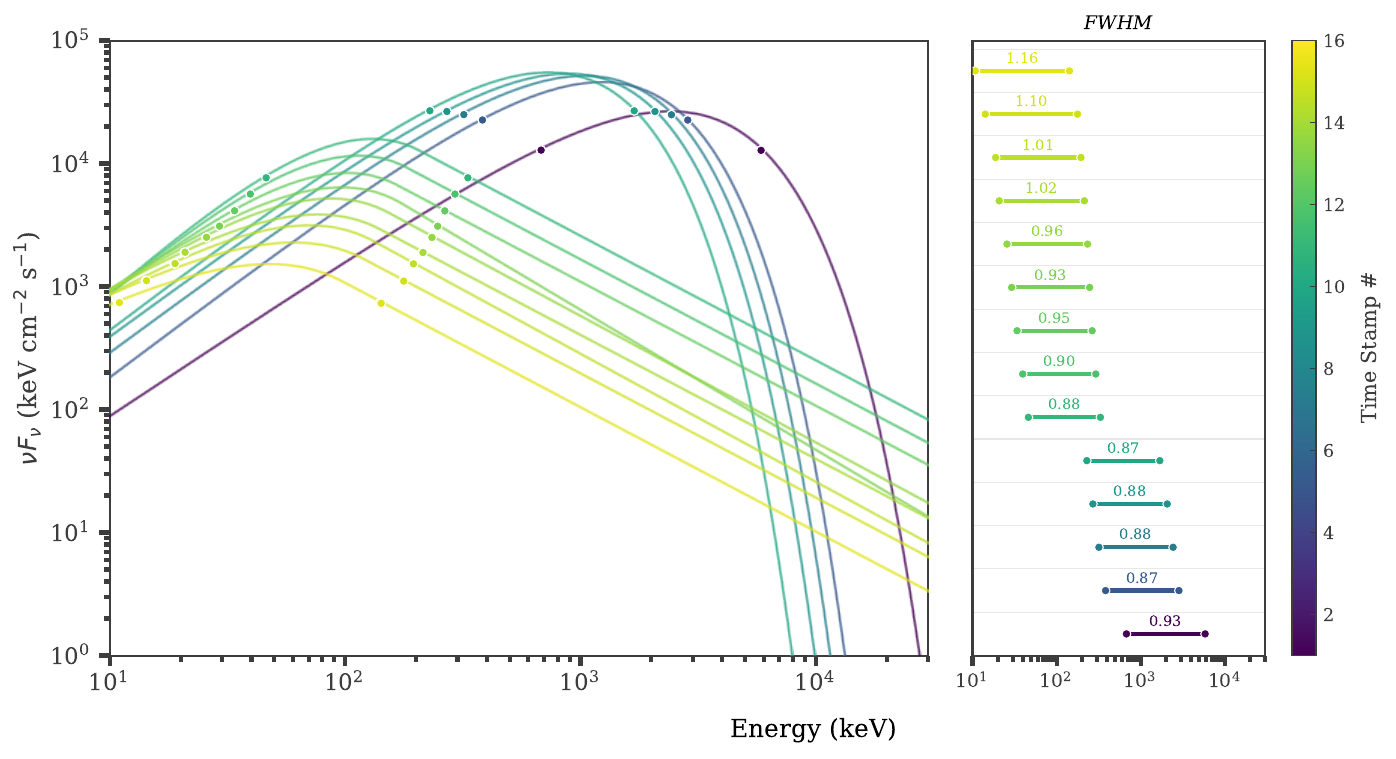}
	\caption{The top and bottom figures represent the \grb and GRB 230812B, respectively. In each figure, the spectral evolution is shown on the left plot and \wid for each bin on the right plot. The color bar indicates the time-bin number. The dots on both the left and right plots mark the $E_{-1/2}$ and $E_{+1/2}$.} 
    \label{fig_spec}
\end{figure}


\appendix

\section{Spectral width of band function}\label{band}
The conventional band function \citep{band1993batse} is described by three parameters, the low-energy spectral index \alp, the high-energy spectral index $\beta$, and the energy $E_0$, which decides the transition between these spectral shapes. Under this representation, the photon number density is expressed as
\begin{align}
	{N(E)= } \left\{
    \begin{array}{ll}
	      A\,E^{\alpha} \exp\left(-\frac{E}{E_0}\right) &\textrm{for}\quad \mbox { $E < \kappa \, E_0$} 	
          \\
        A\,(\kappa\,E_0)^{\kappa} E^{\beta} \exp({-\kappa}) & \textrm{for}\quad \mbox {E$\,\geq\kappa \, E_0$}
    	  \\
    \end{array}	  
	\right.
	\label{eq1}
\end{align}
where, $\kappa = \alpha-\beta$. The peak energy $E_p$ in the $E^2\,N(E)$ representation will 
then be $(\alpha + 2 )E_0$ and the corresponding flux will be 
\begin{align}
    F(E_p) = A\,(\eta E_0)^{\eta} \exp(-\eta)
    \label{eq2}
\end{align}
where, $\eta = \alpha+2$. The spectral width at half maximum flux  is defined as \wid $=\log_{10}(E_{+1/2}/E_{-1/2})$ and $F(E_{+1/2}) \equiv F(E_{-1/2}) = F(E_p)/2$. Here, $E_{-1/2}<E_p$ and $E_{+1/2}>E_p$. Using equations 
\ref{eq1} and \ref{eq2}, we obtain
\begin{align}
    -\frac{E_{-1/2}}{\eta\,E_0}\exp\left(-\frac{E_{-1/2}}{\eta\,E_0}\right) = -\frac{1}{e\,2^{1/\eta}}
    \label{eq3}
\end{align}
or
\begin{align}
    E_{-1/2} = -\eta\,E_0\,\mathscr{W}({-\zeta_\eta})
    \label{eq_e1}
\end{align}
where, $\mathscr{W}$ is the Lambert W function and $\zeta_\eta = (e\,2^{1/\eta})^{-1}$. The positive $E_{-1/2}$ corresponds to the negative solution of $\mathscr{W}$. Again using equations \ref{eq1} and \ref{eq2}, we obtain
\begin{align}
    (\kappa\,E_0)^{\kappa}\,E_{+1/2}^{\beta + 2}\,\exp{(-\kappa)} &= \frac{1}{2}{(\eta\,E_0)^\eta\,\exp{(-\eta)}} \nonumber \\
\end{align}
and hence,
\begin{align}
   E_{+1/2}= \left[\frac{1}{2}\{{\eta^{\eta}\,\kappa^{-\kappa}\,E_0^{\beta + 2}\exp{[-(\beta + 2)]} }\}\right]^{\frac{1}{\beta+2}}
    \label{eq_e2}
\end{align}
Using equations \ref{eq_e1} and \ref{eq_e2}, \wid can be expressed as 
\begin{align}
    \mathcal{W} &= 
    \log\left\{\frac{{(2^{-1}\,\eta^{\eta}\,\kappa^{-\kappa})^{\frac{1}{\beta+2}}\,\exp{(-1)}}}{-\eta\,\mathscr{W}({-\zeta_\eta})}\right\}
    \label{eq_width}
\end{align}
In the reconstructed Band function, $\beta$ is evaluated using equations \ref{eq_width}. Thus, \wid being a free parameter implies
$\beta$ is also a free parameter. Thus, the uncertainties in the \wid will be governed by $\alpha$ as well as $\beta$. 

\section{Spectral width of Power-law with an exponential cutoff function (CPL)} \label{CPL}
The cutoff power law function (CPL) is defined by two parameters, the power-law index p, and the cut-off energy $E_c$, and represented by
\begin{align}
	N(E)= K\,E^p \exp\left(-\frac{E}{E_c}\right)	
    \label{eq1c}
\end{align}
The spectrum in $E^2 N(E)$ representation peaks at energy $E_p (= E_c\,[p+2])$ and the corresponding flux will be 
\begin{align}
    F(E_p) = K\,[(p+2)E_c]^{p+2} \exp[-(p+2)]
    \label{eq2c}
\end{align}
Again, from the definition of \wid (\S xx), we get
\begin{align}
    E_{-1/2} = \frac{2.303(p+2)\mathcal{W}E_c}{10^{\mathcal{W}}-1}
    \label{eq3c}
\end{align}
On substituting equation \ref{eq3c} in $F(E_{-1/2}) = F(E_p)/2$, we obtain the power-law index as a function of \wid, 
\begin{align}
    p = \frac{ln(0.5)}{ln(G)+1-G} -2.0
    \label{eq4c}
\end{align}
and $G = 2.303\, \mathcal{W}/(10^{\mathcal{W}}-1)$.

\section{Evolving fireball model}\label{fireball}
Under the relativistically expanding fireball scenario, the observed emission will be a superposition of multiple Planck functions with temperature varying as $R^{-\xi}$ \citep{peer_08, Gupta_2024}. The composite spectrum at any instant $t$ (corresponding to on-axis radius $R$), will be 
\citep{Gupta_2024}
\begin{align}
	F(E,R) \approx \mathcal{C}R^3E^4 \int\limits_{\beta}^{1} \frac{\mu d\mu}{(1-\beta\mu)^2\left[\exp\left({\frac{E}{kT_{mbb}}}\right)-1\right]}
    \label{eq-mb1}
\end{align}
Here, $\beta c$ is the expansion velocity of thermal plasma, $\mu = \cos\theta$ with $\theta$ being the angle subtended between the line of sight and the off-axis region, and the constant $\mathcal{C} = 4\pi (1-\beta)^2/(h^3c^2D_L^2)$. The temperature, $T_{mbb}$ is given by \citep{Gupta_2024}
\begin{align}
    T_{mbb} = \frac{T_{ph}[R_{ph}(1-\beta\mu)]^\xi(1+\beta\mu)}{[R(1-\beta)]^\xi}
    \label{eq-mb2}
\end{align}
where, $T_{ph}$ is the temperature at on-axis photospheric radius, $R_{ph}$. Since $\beta \approx 1$, the composite spectrum will marginally differ from the Planck function. The dominant contribution in the composite thermal spectrum (for $0\leq\xi\leq2/3$) will be associated with the on-axis emission. 
The peak energy $E_p$ of the composite spectrum will then be governed by the blackbody emission at 
the temperature $T_{mbb,p} = {T_{ph}\,(1+\beta)}(R_{ph}/R)^\xi$ and hence, 
$E_p = 3.93\,k\,{T_{ph}\,(1+\beta)}(R_{ph}/R)^\xi$. The flux at this frequency will be  
\begin{align}
	\mathcal{F}_{E_p}(R) \approx 0.02\, \frac{\mathcal{C}R^3E_p^4}{(1-\beta)^2}
    \label{eq-mb3} 
\end{align}
Since $E_p \propto R^{-\xi}$, when the fireball evolves from $R_0$ to $R$ ($=R_0+  2c\beta\Gamma^2\Delta t$) over the duration $\Delta t$, the $E_p$ shifts towards the low energy. The ratio of peak flux at $R_0$ and $R$ will be
\begin{align}
	\mathcal{H} &= \frac{\mathcal{F}_{E_{p0}}(R_0)}{\mathcal{F}_{E_p}(R)}\nonumber\\
    & = \left(\frac{R_0}{R}\right)^{(3-4\xi)} \nonumber\\
	\label{eq:H}
    & = \left(1-2c\beta\Gamma^2\Delta t/R\right)^{(3-4\xi)}
\end{align}
As $R\rightarrow \infty$, $\mathcal{H}$ approaches unity. The ratio of the peak energies at $R$ and $R_0$
\begin{align}
	\Delta E &= \frac{E_{p0}}{E_p} \nonumber \\
    & = \left(\frac{R_0}{R}\right)^{-\xi} \nonumber\\
	\label{eq:I}
    & = \left(1-2c\beta\Gamma^2\Delta t/R\right)^{-\xi}
\end{align}
and $\Delta E$ approaches unity as $R\rightarrow\infty$. Though it appears that when \rat approaches unity \wid increases; however, $\Delta E$ approaches unity faster, thereby reducing \wid to the instantaneous spectrum.

\section{Synchrotron model}\label{synch}
The evolution of the relativistic Maxwellian distribution losing its energy under the synchrotron process can be described by
\begin{align}
	\frac{\partial N(\gamma,t)}{\partial t} -\frac{\partial}{\partial \gamma}[P(\gamma)N(\gamma,t)] = q(\gamma,t)
    \label{eq_part}
\end{align}
where,
\begin{align}
    q(\gamma,t) = q_0\, \gamma^2\,\exp\left(\frac{-\gamma}{\gamma_c}\right) \delta(t-t_0)
\end{align}
is the electron distribution injected at $t_0$ and $P(\gamma)=B^2\gamma^2/ \mathcal{D}$ is an energy loss rate due to the synchrotron process and $\mathcal{D} = 6\pi mc/\sigma_T$ 
\citep{rybicki}.
The number density at energy $\gamma$ during the instant $t$ to $t+dt$ will be 
\cite{atoyan_99} 
\begin{align}
    N & = \frac{1}{\gamma^2}\int\limits_{t_0}^{t} \xi_\gamma^2(x)\,q[\xi_\gamma(x)]\,\delta(x-t_0) dx\nonumber\\
    & = \frac{q_0\,\gamma^2}{[1-\gamma\,(t-t_0)/\mathcal{D}]^4}\exp\left(\frac{-\xi_{\gamma}(t_0)}{\gamma_c}\right)
    \label{eq_n}
\end{align}
where, 
\begin{align}
    \xi_\gamma(x) = {\gamma}\left\{{1-\left[\frac{\gamma(t-x)}{\mathcal{D}}\right]}\right\}^{-1}
\end{align}
is the energy of the particle at the instant $x$ which will decrease to $\gamma$ at $t$. The synchrotron emissivity due to this evolving electron distribution can be obtained from
\begin{align}
    \mathcal{E}(\nu,t) = \int \limits_{1}^{\infty} N(\gamma,t) P_s(\gamma,\nu) d\gamma
    \label{se1}
\end{align}
where, $P_s(\gamma,\nu)$ is the single particle emissivity \cite{ghisellinni_98}. The \wid of the time-integrated spectrum is obtained numerically for the selected time bins to obtain its evolution.

\bibliography{width}{}

@ARTICLE{Bromberg_090902,
       author = {{Bromberg}, Omer and {Mikolitzky}, Ziv and {Levinson}, Amir},
        title = "{Sub-photospheric Emission from Relativistic Radiation Mediated Shocks in GRBs}",
      journal = {\apj},
         year = 2011,
        month = jun,
       volume = {733},
       number = {2},
          eid = {85},
        pages = {85},
          doi = {10.1088/0004-637X/733/2/85},
archivePrefix = {arXiv},
       eprint = {1101.4232},
 primaryClass = {astro-ph.HE},
       adsurl = {https://ui.adsabs.harvard.edu/abs/2011ApJ...733...85B}
}

@ARTICLE{mizuta_090902,
       author = {{Mizuta}, Akira and {Nagataki}, Shigehiro and {Aoi}, Junichi},
        title = "{Thermal Radiation from Gamma-ray Burst Jets}",
      journal = {\apj},
         year = 2011,
        month = may,
       volume = {732},
       number = {1},
          eid = {26},
        pages = {26},
          doi = {10.1088/0004-637X/732/1/26},
archivePrefix = {arXiv},
       eprint = {1006.2440},
 primaryClass = {astro-ph.HE},
       adsurl = {https://ui.adsabs.harvard.edu/abs/2011ApJ...732...26M}
}

@ARTICLE{zhang_090902,
       author = {{Zhang}, Bin-Bin and {Zhang}, Bing and {Liang}, En-Wei and {Fan}, Yi-Zhong and {Wu}, Xue-Feng and {Pe'er}, Asaf and {Maxham}, Amanda and {Gao}, He and {Dong}, Yun-Ming},
        title = "{A Comprehensive Analysis of Fermi Gamma-ray Burst Data. I. Spectral Components and the Possible Physical Origins of LAT/GBM GRBs}",
      journal = {\apj},
         year = 2011,
        month = apr,
       volume = {730},
       number = {2},
          eid = {141},
        pages = {141},
          doi = {10.1088/0004-637X/730/2/141},
archivePrefix = {arXiv},
       eprint = {1009.3338},
 primaryClass = {astro-ph.HE},
       adsurl = {https://ui.adsabs.harvard.edu/abs/2011ApJ...730..141Z}
}

@article{dynesty,
	author = {Speagle, Joshua S},
    title = {dynesty: a dynamic nested sampling package for estimating Bayesian posteriors and evidences},
    journal = {Monthly Notices of the Royal Astronomical Society},
    volume = {493},
    number = {3},
    pages = {3132-3158},
    year = {2020},
    month = {02},
    issn = {0035-8711},
    doi = {10.1093/mnras/staa278},
    url = {https://doi.org/10.1093/mnras/staa278},
    eprint = {https://academic.oup.com/mnras/article-pdf/493/3/3132/32890730/staa278.pdf},
}

@BOOK{rybicki,
       author = {{Rybicki}, George B. and {Lightman}, Alan P.},
        title = {Radiative processes in astrophysics},
         year = 1979,
       adsurl = {https://ui.adsabs.harvard.edu/abs/1979rpa..book.....R}
}

@ARTICLE{vurm_13,
       author = {{Vurm}, Indrek and {Lyubarsky}, Yuri and {Piran}, Tsvi},
        title = "{On Thermalization in Gamma-Ray Burst Jets and the Peak Energies of Photospheric Spectra}",
      journal = {\apj},
         year = 2013,
        month = feb,
       volume = {764},
       number = {2},
          eid = {143},
        pages = {143},
          doi = {10.1088/0004-637X/764/2/143},
archivePrefix = {arXiv},
       eprint = {1209.0763},
 primaryClass = {astro-ph.HE},
       adsurl = {https://ui.adsabs.harvard.edu/abs/2013ApJ...764..143V}
}

@ARTICLE{yu_bright_fermi,
       author = {{Yu}, Hoi-Fung and {Preece}, Robert D. and {Greiner}, Jochen and {Narayana Bhat}, P. and {Bissaldi}, Elisabetta and {Briggs}, Michael S. and {Cleveland}, William H. and {Connaughton}, Valerie and {Goldstein}, Adam and {von Kienlin}, Andreas and {Kouveliotou}, Chryssa and {Mailyan}, Bagrat and {Meegan}, Charles A. and {Paciesas}, William S. and {Rau}, Arne and {Roberts}, Oliver J. and {Veres}, P{\'e}ter and {Wilson-Hodge}, Colleen and {Zhang}, Bin-Bin and {van Eerten}, Hendrik J.},
        title = "{The Fermi GBM gamma-ray burst time-resolved spectral catalog: brightest bursts in the first four years}",
      journal = {\aap},
         year = 2016,
        month = apr,
       volume = {588},
          eid = {A135},
        pages = {A135},
          doi = {10.1051/0004-6361/201527509},
archivePrefix = {arXiv},
       eprint = {1601.05206},
 primaryClass = {astro-ph.HE},
       adsurl = {https://ui.adsabs.harvard.edu/abs/2016A&A...588A.135Y}
}

@ARTICLE{22_fluence,
       author = {{Malacaria}, C. and {Meegan}, C. and {Fermi GBM Team}},
        title = "{GRB 220426A: Fermi GBM detection}",
      journal = {GRB Coordinates Network},
         year = 2022,
        month = apr,
       volume = {31955},
        pages = {1},
       adsurl = {https://ui.adsabs.harvard.edu/abs/2022GCN.31955....1M}
}

@ARTICLE{23_fluence,
       author = {{Roberts}, O.~J. and {Meegan}, C. and {Lesage}, S. and {Burns}, E. and {Dalessi}, S. and {Fermi GBM Team}},
        title = "{GRB 230812B: Fermi GBM Observation of a very bright burst}",
      journal = {GRB Coordinates Network},
         year = 2023,
        month = aug,
       volume = {34391},
        pages = {1},
       adsurl = {https://ui.adsabs.harvard.edu/abs/2023GCN.34391....1R}
}

@ARTICLE{piran_review_fireball,
       author = {{Piran}, T.},
        title = "{Gamma-ray bursts and the fireball model}",
      journal = {\physrep},
         year = 1999,
        month = jun,
       volume = {314},
       number = {6},
        pages = {575-667},
          doi = {10.1016/S0370-1573(98)00127-6},
archivePrefix = {arXiv},
       eprint = {astro-ph/9810256},
 primaryClass = {astro-ph},
       adsurl = {https://ui.adsabs.harvard.edu/abs/1999PhR...314..575P}
}

@ARTICLE{uhm_zhang_curvature,
       author = {{Uhm}, Z. Lucas and {Zhang}, Bing},
        title = "{On the Curvature Effect of a Relativistic Spherical Shell}",
      journal = {\apj},
         year = 2015,
        month = jul,
       volume = {808},
       number = {1},
          eid = {33},
        pages = {33},
          doi = {10.1088/0004-637X/808/1/33},
archivePrefix = {arXiv},
       eprint = {1411.0118},
 primaryClass = {astro-ph.HE},
       adsurl = {https://ui.adsabs.harvard.edu/abs/2015ApJ...808...33U}
}

@ARTICLE{ruffini_13,
       author = {{Ruffini}, R. and {Siutsou}, I.~A. and {Vereshchagin}, G.~V.},
        title = "{A Theory of Photospheric Emission from Relativistic Outflows}",
      journal = {\apj},
         year = 2013,
        month = jul,
       volume = {772},
       number = {1},
          eid = {11},
        pages = {11},
          doi = {10.1088/0004-637X/772/1/11},
archivePrefix = {arXiv},
       eprint = {1110.0407},
 primaryClass = {astro-ph.CO},
       adsurl = {https://ui.adsabs.harvard.edu/abs/2013ApJ...772...11R}
}

@article{gupta_230307,
	author = {{Gupta, S.} and {Gupta, R.} and {Chattopadhayay, T.} and {Sahayanathan, S.} and {Frederiks, D.} and {Svinkin, D.} and {Bhattacharya, D.} and {Racusin, J.} and {Vadawale, S.} and {Bhalerao, V.} and {Lysenko, A.} and {Ridnaia, A.} and {Tsvetkova, A.} and {Ulanov, M.}},
	title = {Time-resolved spectro-polarimetric analysis of extremely bright GRB 230307A: Possible evidence of evolution from photospheric to synchrotron dominated emission},
	DOI= "10.1051/0004-6361/202555055",
	url= "https://doi.org/10.1051/0004-6361/202555055",
	journal = {Astronomy \& Astrophysics},
	year = 2025,
	volume = 701,
	pages = "A172",
}

@ARTICLE{Burgess_nat,
       author = {{Burgess}, J. Michael and {B{\'e}gu{\'e}}, Damien and {Greiner}, Jochen and {Giannios}, Dimitrios and {Bacelj}, Ana and {Berlato}, Francesco},
        title = "{Gamma-ray bursts as cool synchrotron sources}",
      journal = {Nature Astronomy},
         year = 2020,
        month = feb,
       volume = {4},
        pages = {174-179},
          doi = {10.1038/s41550-019-0911-z},
archivePrefix = {arXiv},
       eprint = {1810.06965},
 primaryClass = {astro-ph.HE},
       adsurl = {https://ui.adsabs.harvard.edu/abs/2020NatAs...4..174B}
}

@ARTICLE{burgess_ryde_2015,
       author = {{Burgess}, J. Michael and {Ryde}, Felix},
        title = "{Are GRB blackbodies an artefact of spectral evolution?}",
      journal = {\mnras},
         year = 2015,
        month = mar,
       volume = {447},
       number = {4},
        pages = {3087-3094},
          doi = {10.1093/mnras/stu2670},
archivePrefix = {arXiv},
       eprint = {1410.7552},
 primaryClass = {astro-ph.HE},
       adsurl = {https://ui.adsabs.harvard.edu/abs/2015MNRAS.447.3087B}
}

@ARTICLE{rees_2005,
       author = {{Rees}, M.~J. and {M{\'e}sz{\'a}ros}, P.},
        title = "{Dissipative Photosphere Models of Gamma-Ray Bursts and X-Ray Flashes}",
      journal = {\apj},
         year = 2005,
        month = aug,
       volume = {628},
       number = {2},
        pages = {847-852},
          doi = {10.1086/430818},
archivePrefix = {arXiv},
       eprint = {astro-ph/0412702},
 primaryClass = {astro-ph},
       adsurl = {https://ui.adsabs.harvard.edu/abs/2005ApJ...628..847R}
}

@ARTICLE{beloborodov_2011,
       author = {{Beloborodov}, Andrei M.},
        title = "{Radiative Transfer in Ultrarelativistic Outflows}",
      journal = {\apj},
         year = 2011,
        month = aug,
       volume = {737},
       number = {2},
          eid = {68},
        pages = {68},
          doi = {10.1088/0004-637X/737/2/68},
archivePrefix = {arXiv},
       eprint = {1011.6005},
 primaryClass = {astro-ph.HE},
       adsurl = {https://ui.adsabs.harvard.edu/abs/2011ApJ...737...68B}
}

@ARTICLE{paczy_1986,
       author = {{Paczynski}, B.},
        title = "{Gamma-ray bursters at cosmological distances}",
      journal = {\apjl},
         year = 1986,
        month = sep,
       volume = {308},
        pages = {L43-L46},
          doi = {10.1086/184740},
       adsurl = {https://ui.adsabs.harvard.edu/abs/1986ApJ...308L..43P}
}

@ARTICLE{Goodman_86,
       author = {{Goodman}, J.},
        title = "{Are gamma-ray bursts optically thick?}",
      journal = {\apjl},
         year = 1986,
        month = sep,
       volume = {308},
        pages = {L47},
          doi = {10.1086/184741},
       adsurl = {https://ui.adsabs.harvard.edu/abs/1986ApJ...308L..47G}
}

@ARTICLE{axelsson_width,
       author = {{Axelsson}, Magnus and {Borgonovo}, Luis},
        title = "{The width of gamma-ray burst spectra}",
      journal = {\mnras},
         year = 2015,
        month = mar,
       volume = {447},
       number = {4},
        pages = {3150-3154},
          doi = {10.1093/mnras/stu2675},
archivePrefix = {arXiv},
       eprint = {1412.5692},
 primaryClass = {astro-ph.HE},
       adsurl = {https://ui.adsabs.harvard.edu/abs/2015MNRAS.447.3150A}
}

@ARTICLE{piran_shemi_naray,
       author = {{Piran}, T. and {Shemi}, A. and {Narayan}, R.},
        title = "{Hydrodynamics of Relativistic Fireballs}",
      journal = {\mnras},
         year = 1993,
        month = aug,
       volume = {263},
        pages = {861},
          doi = {10.1093/mnras/263.4.861},
archivePrefix = {arXiv},
       eprint = {astro-ph/9301004},
 primaryClass = {astro-ph},
       adsurl = {https://ui.adsabs.harvard.edu/abs/1993MNRAS.263..861P}
}

@ARTICLE{burgess_width,
       author = {{Burgess}, J.~M.},
        title = "{Is spectral width a reliable measure of GRB emission physics?}",
      journal = {\aap},
         year = 2019,
        month = sep,
       volume = {629},
          eid = {A69},
        pages = {A69},
          doi = {10.1051/0004-6361/201935140},
archivePrefix = {arXiv},
       eprint = {1705.05718},
 primaryClass = {astro-ph.HE},
       adsurl = {https://ui.adsabs.harvard.edu/abs/2019A&A...629A..69B}
}

@ARTICLE{rees94,
   author = {{Rees}, M.~J. and {Meszaros}, P.},
    title = "{Unsteady outflow models for cosmological gamma-ray bursts}",
  journal = {\apjl},
   eprint = {astro-ph/9404038},
     year = 1994,
    month = aug,
   volume = 430,
    pages = {L93-L96},
      doi = {10.1086/187446},
   adsurl = {http://adsabs.harvard.edu/abs/1994ApJ...430L..93R}
}

@ARTICLE{cohen_1997,
       author = {{Cohen}, Ehud and {Piran}, Tsvi},
        title = "{The Implications of Direct Redshift Measurement of Gamma-Ray Bursts}",
      journal = {\apjl},
         year = 1997,
        month = oct,
       volume = {488},
       number = {1},
        pages = {L7-L10},
          doi = {10.1086/310916},
archivePrefix = {arXiv},
       eprint = {astro-ph/9706045},
 primaryClass = {astro-ph},
       adsurl = {https://ui.adsabs.harvard.edu/abs/1997ApJ...488L...7C}
}

@ARTICLE{sari_1997,
       author = {{Sari}, Re'em and {Piran}, Tsvi},
        title = "{Cosmological gamma-ray bursts: internal versus external shocks}",
      journal = {\mnras},
         year = 1997,
        month = may,
       volume = {287},
       number = {1},
        pages = {110-116},
          doi = {10.1093/mnras/287.1.110},
archivePrefix = {arXiv},
       eprint = {astro-ph/9608152},
 primaryClass = {astro-ph},
       adsurl = {https://ui.adsabs.harvard.edu/abs/1997MNRAS.287..110S}
}

@ARTICLE{kobayashi_1997,
       author = {{Kobayashi}, Shiho and {Piran}, Tsvi and {Sari}, Re'em},
        title = "{Can Internal Shocks Produce the Variability in Gamma-Ray Bursts?}",
      journal = {\apj},
         year = 1997,
        month = nov,
       volume = {490},
        pages = {92},
          doi = {10.1086/512791},
archivePrefix = {arXiv},
       eprint = {astro-ph/9705013},
 primaryClass = {astro-ph},
       adsurl = {https://ui.adsabs.harvard.edu/abs/1997ApJ...490...92K}
}

@ARTICLE{Rees_internal,
       author = {{Rees}, M.~J.},
        title = "{The M87 jet: internal shocks in a plasma beam?}",
      journal = {\mnras},
         year = 1978,
        month = sep,
       volume = {184},
        pages = {61P-65P},
          doi = {10.1093/mnras/184.1.61P},
       adsurl = {https://ui.adsabs.harvard.edu/abs/1978MNRAS.184P..61R}
}

@ARTICLE{spada2001,
       author = {{Spada}, Maddalena and {Ghisellini}, Gabriele and {Lazzati}, Davide and {Celotti}, Annalisa},
        title = "{Internal shocks in the jets of radio-loud quasars}",
      journal = {\mnras},
         year = 2001,
        month = aug,
       volume = {325},
       number = {4},
        pages = {1559-1570},
          doi = {10.1046/j.1365-8711.2001.04557.x},
archivePrefix = {arXiv},
       eprint = {astro-ph/0103424},
 primaryClass = {astro-ph},
       adsurl = {https://ui.adsabs.harvard.edu/abs/2001MNRAS.325.1559S}
}

@ARTICLE{daigne_1998,
       author = {{Daigne}, F. and {Mochkovitch}, R.},
        title = "{Gamma-ray bursts from internal shocks in a relativistic wind: temporal and spectral properties}",
      journal = {\mnras},
         year = 1998,
        month = may,
       volume = {296},
       number = {2},
        pages = {275-286},
          doi = {10.1046/j.1365-8711.1998.01305.x},
archivePrefix = {arXiv},
       eprint = {astro-ph/9801245},
 primaryClass = {astro-ph},
       adsurl = {https://ui.adsabs.harvard.edu/abs/1998MNRAS.296..275D}
}

@ARTICLE{daigne_2011,
       author = {{Daigne}, F. and {Bo{\v{s}}njak}, {\v{Z}}. and {Dubus}, G.},
        title = "{Reconciling observed gamma-ray burst prompt spectra with synchrotron radiation?}",
      journal = {\aap},
         year = 2011,
        month = feb,
       volume = {526},
          eid = {A110},
        pages = {A110},
          doi = {10.1051/0004-6361/201015457},
archivePrefix = {arXiv},
       eprint = {1009.2636},
 primaryClass = {astro-ph.HE},
       adsurl = {https://ui.adsabs.harvard.edu/abs/2011A&A...526A.110D}
}

@ARTICLE{bbzhang_2016,
       author = {{Zhang}, Bin-Bin and {Uhm}, Z. Lucas and {Connaughton}, Valerie and {Briggs}, Michael S. and {Zhang}, Bing},
        title = "{Synchrotron Origin of the Typical GRB Band Function{\textemdash}A Case Study of GRB 130606B}",
      journal = {\apj},
         year = 2016,
        month = jan,
       volume = {816},
       number = {2},
          eid = {72},
        pages = {72},
          doi = {10.3847/0004-637X/816/2/72},
archivePrefix = {arXiv},
       eprint = {1505.05858},
 primaryClass = {astro-ph.HE},
       adsurl = {https://ui.adsabs.harvard.edu/abs/2016ApJ...816...72Z}
}

@ARTICLE{Bonjak2009,
       author = {{Bo{\v{s}}njak}, {\v{Z}}. and {Daigne}, F. and {Dubus}, G.},
        title = "{Prompt high-energy emission from gamma-ray bursts in the internal shock model}",
      journal = {\aap},
         year = 2009,
        month = may,
       volume = {498},
       number = {3},
        pages = {677-703},
          doi = {10.1051/0004-6361/200811375},
archivePrefix = {arXiv},
       eprint = {0811.2956},
 primaryClass = {astro-ph},
       adsurl = {https://ui.adsabs.harvard.edu/abs/2009A&A...498..677B}
}

@ARTICLE{peer_08,
       author = {{Pe'er}, Asaf},
        title = "{Temporal Evolution of Thermal Emission from Relativistically Expanding Plasma}",
      journal = {\apj},
         year = 2008,
        month = jul,
       volume = {682},
       number = {1},
        pages = {463-473},
          doi = {10.1086/588136},
archivePrefix = {arXiv},
       eprint = {0802.0725},
 primaryClass = {astro-ph},
       adsurl = {https://ui.adsabs.harvard.edu/abs/2008ApJ...682..463P}
}

@ARTICLE{uhm_14,
       author = {{Uhm}, Z. Lucas and {Zhang}, Bing},
        title = "{Fast-cooling synchrotron radiation in a decaying magnetic field and {\ensuremath{\gamma}}-ray burst emission mechanism}",
      journal = {Nature Physics},
         year = 2014,
        month = may,
       volume = {10},
       number = {5},
        pages = {351-356},
          doi = {10.1038/nphys2932},
archivePrefix = {arXiv},
       eprint = {1303.2704},
 primaryClass = {astro-ph.HE},
       adsurl = {https://ui.adsabs.harvard.edu/abs/2014NatPh..10..351U}
}

@ARTICLE{lundman2013,
       author = {{Lundman}, C. and {Pe'er}, A. and {Ryde}, F.},
        title = "{A theory of photospheric emission from relativistic, collimated outflows}",
      journal = {\mnras},
         year = 2013,
        month = jan,
       volume = {428},
       number = {3},
        pages = {2430-2442},
          doi = {10.1093/mnras/sts219},
archivePrefix = {arXiv},
       eprint = {1208.2965},
 primaryClass = {astro-ph.HE},
       adsurl = {https://ui.adsabs.harvard.edu/abs/2013MNRAS.428.2430L}
}

@ARTICLE{peer090902B,
       author = {{Pe'Er}, Asaf and {Zhang}, Bin-Bin and {Ryde}, Felix and {McGlynn}, Sin{\'e}ad and {Zhang}, Bing and {Preece}, Robert D. and {Kouveliotou}, Chryssa},
        title = "{The connection between thermal and non-thermal emission in gamma-ray bursts: general considerations and GRB 090902B as a case study}",
      journal = {\mnras},
         year = 2012,
        month = feb,
       volume = {420},
       number = {1},
        pages = {468-482},
          doi = {10.1111/j.1365-2966.2011.20052.x},
archivePrefix = {arXiv},
       eprint = {1007.2228},
 primaryClass = {astro-ph.HE},
       adsurl = {https://ui.adsabs.harvard.edu/abs/2012MNRAS.420..468P}
}

@ARTICLE{zhang_icmart,
       author = {{Zhang}, Bing and {Yan}, Huirong},
        title = "{The Internal-collision-induced Magnetic Reconnection and Turbulence (ICMART) Model of Gamma-ray Bursts}",
      journal = {\apj},
         year = 2011,
        month = jan,
       volume = {726},
       number = {2},
          eid = {90},
        pages = {90},
          doi = {10.1088/0004-637X/726/2/90},
archivePrefix = {arXiv},
       eprint = {1011.1197},
 primaryClass = {astro-ph.HE},
       adsurl = {https://ui.adsabs.harvard.edu/abs/2011ApJ...726...90Z}
}

@ARTICLE{meszaros93,
   author = {{Meszaros}, P. and {Rees}, M.~J.},
    title = "{Relativistic fireballs and their impact on external matter - Models for cosmological gamma-ray bursts}",
  journal = {\apj},
     year = 1993,
    month = mar,
   volume = 405,
    pages = {278-284},
      doi = {10.1086/172360},
   adsurl = {http://adsabs.harvard.edu/abs/1993ApJ...405..278M}
}

@ARTICLE{34694,
       author = {{Roberts}, O.~J. and {Cleveland}, W. and {Fermi GBM Team}},
        title = "{GRB 230812B: Update on Bad Time Intervals for Fermi GBM data}",
      journal = {GRB Coordinates Network},
         year = 2024,
        month = feb,
       volume = {35660},
        pages = {1},
       adsurl = {https://ui.adsabs.harvard.edu/abs/2024GCN.35660....1R}
}

@ARTICLE{burgess_ryde_yu,
       author = {{Burgess}, J. Michael and {Ryde}, Felix and {Yu}, Hoi-Fung},
        title = "{Taking the band function too far: a tale of two {\ensuremath{\alpha}}'s}",
      journal = {\mnras},
         year = 2015,
        month = aug,
       volume = {451},
       number = {2},
        pages = {1511-1521},
          doi = {10.1093/mnras/stv775},
archivePrefix = {arXiv},
       eprint = {1410.7647},
 primaryClass = {astro-ph.HE},
       adsurl = {https://ui.adsabs.harvard.edu/abs/2015MNRAS.451.1511B}
}

@ARTICLE{Peer_2006,
       author = {{Pe'er}, Asaf and {Zhang}, Bing},
        title = "{Synchrotron Emission in Small-Scale Magnetic Fields as a Possible Explanation for Prompt Emission Spectra of Gamma-Ray Bursts}",
      journal = {\apj},
         year = 2006,
        month = dec,
       volume = {653},
       number = {1},
        pages = {454-461},
          doi = {10.1086/508681},
archivePrefix = {arXiv},
       eprint = {astro-ph/0605641},
 primaryClass = {astro-ph},
       adsurl = {https://ui.adsabs.harvard.edu/abs/2006ApJ...653..454P}
}

@ARTICLE{sari98,
   author = {{Sari}, R. and {Piran}, T. and {Narayan}, R.},
    title = "{Spectra and Light Curves of Gamma-Ray Burst Afterglows}",
  journal = {\apjl},
   eprint = {astro-ph/9712005},
     year = 1998,
    month = apr,
   volume = 497,
    pages = {L17-L20},
      doi = {10.1086/311269},
   adsurl = {http://adsabs.harvard.edu/abs/1998ApJ...497L..17S}
}

@article{scargle1998studies,
  title={Studies in astronomical time series analysis. V. Bayesian blocks, a new method to analyze structure in photon counting data},
  author={Scargle, Jeffrey D},
  journal={The Astrophysical Journal},
  volume={504},
  number={1},
  pages={405},
  year={1998},
  publisher={IOP Publishing}
}

@ARTICLE{Sharma_etal_2019,
       author = {{Sharma}, Vidushi and {Iyyani}, Shabnam and {Bhattacharya}, Dipankar and
         {Chattopadhyay}, Tanmoy and {Rao}, A.~R. and {Aarthy}, E. and
         {Vadawale}, Santosh V. and {Mithun}, N.~P.~S. and
         {Bhalerao}, Varun. B. and {Ryde}, Felix and {Pe'er}, Asaf},
        title = "{Time-varying Polarized Gamma-Rays from GRB 160821A: Evidence for Ordered Magnetic Fields}",
      journal = {\apjl},
         year = 2019,
        month = sep,
       volume = {882},
       number = {1},
          eid = {L10},
        pages = {L10},
          doi = {10.3847/2041-8213/ab3a48},
archivePrefix = {arXiv},
       eprint = {1908.10885},
 primaryClass = {astro-ph.HE},
       adsurl = {https://ui.adsabs.harvard.edu/abs/2019ApJ...882L..10S}
}

@ARTICLE{GillandGranot2021,
       author = {{Gill}, Ramandeep and {Granot}, Jonathan},
        title = "{Temporal evolution of prompt GRB polarization}",
      journal = {\mnras},
         year = 2021,
        month = jun,
       volume = {504},
       number = {2},
        pages = {1939-1958},
          doi = {10.1093/mnras/stab1013},
archivePrefix = {arXiv},
       eprint = {2101.06777},
 primaryClass = {astro-ph.HE},
       adsurl = {https://ui.adsabs.harvard.edu/abs/2021MNRAS.504.1939G}
}

@ARTICLE{Preece_etal_1998,
       author = {{Preece}, R.~D. and {Briggs}, M.~S. and {Mallozzi}, R.~S. and {Pendleton}, G.~N. and {Paciesas}, W.~S. and {Band}, D.~L.},
        title = "{The Synchrotron Shock Model Confronts a ``Line of Death'' in the BATSE Gamma-Ray Burst Data}",
      journal = {\apjl},
         year = 1998,
        month = oct,
       volume = {506},
       number = {1},
        pages = {L23-L26},
          doi = {10.1086/311644},
archivePrefix = {arXiv},
       eprint = {astro-ph/9808184},
 primaryClass = {astro-ph},
       adsurl = {https://ui.adsabs.harvard.edu/abs/1998ApJ...506L..23P}
}

@ARTICLE{Iyyani_etal_2016,
       author = {{Iyyani}, S. and {Ryde}, F. and {Burgess}, J.~M. and {Pe'er}, A. and {B{\'e}gu{\'e}}, D.},
        title = "{Synchrotron emission in GRBs observed by Fermi: its limitations and the role of the photosphere}",
      journal = {\mnras},
         year = 2016,
        month = feb,
       volume = {456},
       number = {2},
        pages = {2157-2171},
          doi = {10.1093/mnras/stv2751},
archivePrefix = {arXiv},
       eprint = {1511.07390},
 primaryClass = {astro-ph.HE},
       adsurl = {https://ui.adsabs.harvard.edu/abs/2016MNRAS.456.2157I}
}

@article{band1993batse,
 title        = {{BATSE Observations of Gamma-Ray Burst Spectra. I. Spectral Diversity}},
 author       = {{Band}, D. and {Matteson}, J. and {Ford}, L. and {Schaefer}, B. and {Palmer}, D. and {Teegarden}, B. and {Cline}, T. and {Briggs}, M. and {Paciesas}, W. and {Pendleton}, G. and {Fishman}, G. and {Kouveliotou}, C. and {Meegan}, C. and {Wilson}, R. and {Lestrade}, P.},
 year         = 1993,
 month        = aug,
 journal      = {ApJ},
 volume       = 413,
 pages        = 281,
 doi          = {10.1086/172995},
 adsurl       = {https://ui.adsabs.harvard.edu/abs/1993ApJ...413..281B}
}

@ARTICLE{Tavani1996,
       author = {{Tavani}, M.},
        title = "{A Shock Emission Model for Gamma-Ray Bursts. II. Spectral Properties}",
      journal = {\apj},
         year = 1996,
        month = aug,
       volume = {466},
        pages = {768},
          doi = {10.1086/177551},
       adsurl = {https://ui.adsabs.harvard.edu/abs/1996ApJ...466..768T}
}

@ARTICLE{crider_1997,
       author = {{Crider}, A. and {Liang}, E.~P. and {Smith}, I.~A. and {Preece}, R.~D. and {Briggs}, M.~S. and {Pendleton}, G.~N. and {Paciesas}, W.~S. and {Band}, D.~L. and {Matteson}, J.~L.},
        title = "{Evolution of the Low-Energy Photon Spectral in Gamma-Ray Bursts}",
      journal = {\apjl},
         year = 1997,
        month = apr,
       volume = {479},
       number = {1},
        pages = {L39-L42},
          doi = {10.1086/310574},
archivePrefix = {arXiv},
       eprint = {astro-ph/9612118},
 primaryClass = {astro-ph},
       adsurl = {https://ui.adsabs.harvard.edu/abs/1997ApJ...479L..39C}
}

@ARTICLE{deng2014,
       author = {{Deng}, Wei and {Zhang}, Bing},
        title = "{Low Energy Spectral Index and E$_{p}$ Evolution of Quasi-thermal Photosphere Emission of Gamma-Ray Bursts}",
      journal = {\apj},
         year = 2014,
        month = apr,
       volume = {785},
       number = {2},
          eid = {112},
        pages = {112},
          doi = {10.1088/0004-637X/785/2/112},
archivePrefix = {arXiv},
       eprint = {1402.5364},
 primaryClass = {astro-ph.HE},
       adsurl = {https://ui.adsabs.harvard.edu/abs/2014ApJ...785..112D}
}

@article{Gupta_2024,
doi = {10.3847/2041-8213/ad5e1e},
url = {https://dx.doi.org/10.3847/2041-8213/ad5e1e},
year = {2024},
month = {jul},
publisher = {The American Astronomical Society},
volume = {970},
number = {1},
pages = {L12},
author = {Gupta, Soumya and Sahayanathan, Sunder},
title = {Investigation of the Gamma-Ray Bursts Prompt Emission Under the Relativistically Expanding Fireball Scenario},
journal = {The Astrophysical Journal Letters}
}

@ARTICLE{220426_fermi,
       author = {{Deng}, Li-Tao and {Lin}, Da-Bin and {Zhou}, Li and {Wang}, Kai and {Yang}, Xing and {Hou}, Shu-Jin and {Li}, Jing and {Wang}, Xiang-Gao and {Lu}, Rui-Jing and {Liang}, En-Wei},
        title = "{Fermi Observations of GRB 220426A: a burst similar to GRB 090902B}",
      journal = {arXiv e-prints},
         year = 2022,
        month = may,
          eid = {arXiv:2205.08737},
        pages = {arXiv:2205.08737},
          doi = {10.48550/arXiv.2205.08737},
archivePrefix = {arXiv},
       eprint = {2205.08737},
 primaryClass = {astro-ph.HE},
       adsurl = {https://ui.adsabs.harvard.edu/abs/2022arXiv220508737D}
}

@ARTICLE{atoyan_99,
       author = {{Atoyan}, A.~M. and {Aharonian}, F.~A.},
        title = "{Modelling of the non-thermal flares in the Galactic microquasar GRS 1915+105}",
      journal = {\mnras},
         year = 1999,
        month = jan,
       volume = {302},
       number = {2},
        pages = {253-276},
          doi = {10.1046/j.1365-8711.1999.02172.x},
       adsurl = {https://ui.adsabs.harvard.edu/abs/1999MNRAS.302..253A}
}

@ARTICLE{ghisellinni_98,
       author = {{Chiaberge}, Marco and {Ghisellini}, Gabriele},
        title = "{Rapid variability in the synchrotron self-Compton model for blazars}",
      journal = {\mnras},
         year = 1999,
        month = jul,
       volume = {306},
       number = {3},
        pages = {551-560},
          doi = {10.1046/j.1365-8711.1999.02538.x},
archivePrefix = {arXiv},
       eprint = {astro-ph/9810263},
 primaryClass = {astro-ph},
       adsurl = {https://ui.adsabs.harvard.edu/abs/1999MNRAS.306..551C}
}

@ARTICLE{zhang_2018,
       author = {{Zhang}, B. -B. and {Zhang}, B. and {Castro-Tirado}, A.~J. and {Dai}, Z.~G. and {Tam}, P. -H.~T. and {Wang}, X. -Y. and {Hu}, Y. -D. and {Karpov}, S. and {Pozanenko}, A. and {Zhang}, F. -W. and {Mazaeva}, E. and {Minaev}, P. and {Volnova}, A. and {Oates}, S. and {Gao}, H. and {Wu}, X. -F. and {Shao}, L. and {Tang}, Q. -W. and {Beskin}, G. and {Biryukov}, A. and {Bondar}, S. and {Ivanov}, E. and {Katkova}, E. and {Orekhova}, N. and {Perkov}, A. and {Sasyuk}, V. and {Mankiewicz}, L. and {{\.Z}arnecki}, A.~F. and {Cwiek}, A. and {Opiela}, R. and {Zadro{\.Z}ny}, A. and {Aptekar}, R. and {Frederiks}, D. and {Svinkin}, D. and {Kusakin}, A. and {Inasaridze}, R. and {Burhonov}, O. and {Rumyantsev}, V. and {Klunko}, E. and {Moskvitin}, A. and {Fatkhullin}, T. and {Sokolov}, V.~V. and {Valeev}, A.~F. and {Jeong}, S. and {Park}, I.~H. and {Caballero-Garc{\'\i}a}, M.~D. and {Cunniffe}, R. and {Tello}, J.~C. and {Ferrero}, P. and {Pandey}, S.~B. and {Jel{\'\i}nek}, M. and {Peng}, F.~K. and {S{\'a}nchez-Ram{\'\i}rez}, R. and {Castell{\'o}n}, A.},
        title = "{Transition from fireball to Poynting-flux-dominated outflow in the three-episode GRB 160625B}",
      journal = {Nature Astronomy},
         year = 2018,
        month = nov,
       volume = {2},
        pages = {69-75},
          doi = {10.1038/s41550-017-0309-8},
archivePrefix = {arXiv},
       eprint = {1612.03089},
 primaryClass = {astro-ph.HE},
       adsurl = {https://ui.adsabs.harvard.edu/abs/2018NatAs...2...69Z}
}

@ARTICLE{chen_21,
       author = {{Chen}, Jia-Ming and {Peng}, Zhao-Yang and {Du}, Tan-Tan and {Yin}, Yue and {Wu}, Hui},
        title = "{GRB 180720B: A GRB with Interesting Spectral Characteristics}",
      journal = {\apj},
         year = 2021,
        month = oct,
       volume = {920},
       number = {1},
          eid = {53},
        pages = {53},
          doi = {10.3847/1538-4357/ac14b8},
       adsurl = {https://ui.adsabs.harvard.edu/abs/2021ApJ...920...53C}
}
\bibliographystyle{aasjournal}
\end{document}